# Natural radioelement concentration in the Troodos Ophiolite Complex of Cyprus

**Michalis Tzortzis and Haralabos Tsertos**[*]

*Department of Physics, University of Cyprus, P.O. Box 20537, 1678 Nicosia, Cyprus.*

**Abstract**

High-resolution γ–ray spectrometry was exploited to determine naturally occurring thorium (Th), uranium (U) and potassium (K) elemental concentrations in the whole area covered by the Troodos Ophiolite Complex of Cyprus. For that purpose, a total of 59 samples from surface soils and 10 from the main rock formations of the region of interest were analysed. Elemental concentrations were determined for Th (range from $2.5 \times 10^{-3}$ *ppm* to 2.0 *ppm*), U (from $8.1 \times 10^{-4}$ *ppm* to 0.6 *ppm*), and K (from $1.3 \times 10^{-4}$ *%* to 1.0 *%*). The average values (A.M ± S.D.) derived are (0.24 ± 0.34) *ppm*, (0.10 ± 0.10) *ppm* and (0.21 ± 0.24) *%*, for Th, U, and K, respectively, in the soils, and (0.52 ± 0.17) *ppm*, (0.17 ± 0.11) *ppm* and (0.49 ± 0.87) *%* in the rocks. From these values, a radioactivity (radioelement) loss of nearly 50% is estimated in the underlying surface soils due to bleaching and eluviation during weathering of the rocks. The measured Th/U ratio exhibits values between 2 and 4, whereas the K/Th ratio is highly variable ranging between $1.5 \times 10^{3}$ and $3.0 \times 10^{4}$.

---

[*] **Corresponding author.** E-mail address: *tsertos@ucy.ac.cy*, Fax: *+357-22892821*.

# Introduction

The main external source of irradiation to the human body is represented by the gamma radiation emitted by naturally occurring radioisotopes, also called terrestrial environmental radiation. These radioisotopes, such as $^{40}$K and the radionuclides from the $^{232}$Th and $^{238}$U series and their decay products, exist at trace levels in all ground formations. Therefore, natural environmental radioactivity and the associated external exposure due to gamma radiation depend primarily on the geological and geographical conditions, and appear at different levels in the soils of each different geological region (UNSCEAR 2000 Report, and further references cited therein). The specific levels of terrestrial environmental radiation are related to the geological composition of each lithologically separated area, and to the content in thorium (Th), uranium (U) and potassium (K) of the rock from which the soils originate in each area.

The island of Cyprus is located in the eastern basin of the Mediterranean Sea and extends to an area of about 9,300 $km^2$. Its characteristic geological formations can be classified into two broad main categories: those that belong to an ophiolite complex (extending to an area of about 3000 $km^2$) and those of sedimentary origin (Figure 1). The Cyprus ophiolite is one of the best preserved and most intensively studied ophiolite complexes in the world, and is known as the Troodos Massif or Troodos Ophiolite Complex (Moores and Vine, 1971; Robinson and Malpas, 1998).

The large geological significance of the Late Cretaceous Troodos ophiolite is focused on its complete and non-disruptive sequence tougher with a displaced slab of altered ultrabasic and basic plutonic complex, stratigraphically overlain successfully by a sheeted dyke complex, extrusive sequence and pelagic sediments (Moores and Vine, 1971; Gass, 1980). Troodos is believed to have formed at several spreading axes in a supra-subduction zone environment as a result of the collision of the African and Eurasian Plates in the Late Cretaceous (Robinson and Malpas, 1998). Geological conditions in the area maintained the whole ophiolite stratigraphy unimpaired through centuries and, hence, both the Troodos Ophiolite Complex and the ocean crust have exactly the same layer structure (Figure 2). Due to subsequent erosion, the primary upward succession of ophiolite stratigraphy was arranged in an outward succession from centrally exposed plutonic rocks to a sheeted dyke complex and peripheral pillow lavas (De Silva, 1997). Out of the complete ophiolite sequence, the volcanic series is divided into the Upper Pillow Lavas (UPL), the Lower Pillow Lavas (LPL), and the Basalt Group (BG), based on colour, mineralogy, abundance of dykes, and the relationship to the sulphide orebodies (Wilson and Ingham, 1959). A simplistic diagram of the ophiolite stratigraphy along with the other formations appearing out of the ophiolite sequence is presented in Figure 3. Generally, in respect to oxide abundance, the ophiolite of Troodos is rich in *CaO* and poor in *$SiO_2$*. Its rocks are mostly iron- and magnesium-saturated or oversaturated, and this allows their characterisation as basic and ultrabasic. None of these rock types belongs to the category of silica-oversaturated, which usually is associated with high Th and U elemental concentrations (Faure, 1986).

This study is part of a general project called "radioisotopes" that had the objective to systematically measure the terrestrial gamma radiation in the island, and to determine its contribution to the annual effective dose equivalent to the population (Tzortzis et al., 2003, 2004). After the completion of the data processing and analysis, it was realised that samples originating from the ophiolitic part of the island exhibited very low activity and elemental concentrations compared to the corresponding results from the samples of sedimentary origin. In this paper, the results from this study regarding Th, U and K elemental concentrations in the whole area of the Troodos ophiolite complex are presented. Although intensively studied regarding many geological aspects, as mentioned before, the Troodos Ophiolite has remained unexplored up to now with respect to natural environmental radioactivity and associated Th, U, and K elemental concentrations appearing in its formations. The present study aims at filling this vacancy in the literature and, more generally, to provide data on Th, U, and K elemental concentration that are also relevant to the ocean crust and mantle of the earth. The experiments have been carried out in the Nuclear Physics Laboratory of the Department of Physics, University of Cyprus, using a high-resolution $\gamma$–ray spectrometry system.

## Materials and methods

### Sample collection, preparation and counting

A total of 115 surface soil samples have been collected throughout the whole accessible area of the island. This area was divided into seven different geological regions, three of which belong to the Troodos ophiolite, as indicated in the simplified map of Figure 1. Samples were collected from different locations falling within the boundaries of each of the seven geological regions considered. More than half (59) of the samples were collected from the Troodos Ophiolite (Regions 4, 6, and 7). In addition, 10 samples from the predominant rock formations of the study area were also collected and pulverised. Collected soil samples were air-dried, sieved through a fine mesh, hermetically sealed in standard 1000-*ml* plastic Marinelli beakers, dry-weighed and stored for about four weeks prior to counting. The exact procedure followed for the sample collection, preparation and counting is analytically described elsewhere (Tzortzis et al., 2003, 2004).

Each plastic sample container was placed on the active area of a shielded High-Purity Germanium (HPGe) detector, and measured for a counting time of 18 *hours*. The energy spectrum of the emitted gamma rays was obtained with a relative photopeak efficiency of ~33% and with an energy resolution (FWHM) of 1.8 *keV* at the 1.33 *MeV* reference transition of $^{60}$Co (details are given in Tzortzis et al., 2003). Depending on the peak background, the Minimum Detectable Activity (MDA) was calculated to be $1.0 \times 10^{-2}$ *Bq kg$^{-1}$* for both $^{232}$Th and $^{238}$U, and $4.0 \times 10^{-2}$ *Bq kg$^{-1}$* for $^{40}$K, for the counting time of 18 *hours*. From these values, detection limits of $2.5 \times 10^{-3}$ *ppm*, $8.1 \times 10^{-4}$ *ppm* and $1.3 \times 10^{-4}$ *%* were derived, for Th, U and K elemental concentrations, respectively.

Following the spectrum analysis, count rates for each detected photopeak and activity per mass unit (specific activity) for each of the detected nuclides are calculated in units of *Bq kg$^{-1}$*. Specific activity of $^{232}$Th, $^{238}$U and $^{40}$K was then converted into total elemental concentration of Th, U and K, respectively. They are reported in units of *ppm* for Th and U, and as a percentage (*%*) for K (Tzortzis et al., 2003).

**Results and discussion**

Samples of ophiolitic origin are categorised in three groups, one for each different geological region as indicated in Figure 1 (Regions 4, 6, and 7), and the average activity and elemental concentrations of Th, U, and K for all the three regions are summarised in Table 1, from soil as well as from rock samples. Such a number of collected samples can be considered as a representative inventory of the various outcropping geological formations in the region of interest.

In general, the calculated elemental concentrations of Th, U and K are rather low in all the samples measured; this is particularly observable in samples collected in Region 4, which exhibits the lowest concentration in all the radionuclides investigated. This region consists of the upper levels of plutonics with their complex structure being well exposed: plagiogranites, which are apparently the result of extreme fractionation of basaltic magma, and the essential nature of layered gabbros, dunite/harzburgite relationships, wehrlite, and formation of chromitite pods (Keller, 1999). On the other hand, Region 7 that is mainly composed of rocks of volcanic sequence, thus the Upper and Lower Pillow Lavas and the Basalt Group, shows the highest activity concentration in all the radionuclides investigated. Concentration of Th ranged from $2.5 \times 10^{-3}$ *ppm* to 2.0 *ppm*, of U from $8.1 \times 10^{-4}$ *ppm* to 0.6 *ppm* and of K from $1.3 \times 10^{-4}$ *%* to 1.0 *%* for all the samples measured. It should be noted that the lower values given correspond to the detection limit (MDA) derived for each of the three radionuclides, since some samples exhibited nearly zero net activity after the subtraction of the laboratory ambient background. Arithmetic mean elemental concentrations over all the samples of ophiolitic origin are (0.24 ± 0.34) *ppm*, (0.10 ± 0.10) *ppm*, (0.21 ± 0.24) *%,* for Th, U and K, respectively, while the island's average values from ground formations are (1.2 ± 1.7) *ppm*, (0.6 ± 0.7) *ppm* and (0.4 ± 0.3) *%,* respectively (Tzortzis and Tsertos, 2004). This indicates that the values of the elemental concentrations, calculated for the samples of ophiolitic origin, are significantly lower than the corresponding ones collected from all over the island.

Th, U, and K elemental concentration ranges derived from investigations conducted in some other known ophiolite complexes of neighbouring regions close and around the Mediterranean Sea are illustrated in Table 2. From this comparison, it can be concluded that the corresponding Th, U, and K elemental concentration values obtained from the present study fall within the range of the reported values from other ophiolites. These results confirm a general trend that the content of those ophiolite formations in radioelements is very low compared to the content of other rock types, e.g of sedimentary origin. In other words, since natural radioactivity is directly related to the soil content in Th, U, and K, the Troodos Ophiolite, and the whole island of Cyprus in general, could be considered as one of the world areas that exhibit very low levels of natural radioactivity. It is also important to underline that this feature reflects a difference in the rock geology of the island of Cyprus compared to that of the neighbouring Mediterranean countries. As mentioned before, the rock types appearing in Cyprus have their origin in a well-preserved ophiolite, and the data derived regarding Th, U, and K elemental concentrations are characteristic for the crust and upper mantle of the earth (Faure, 1986).

Th/U and K/Th correlation plots are drawn in Figure 4a,b, which may provide information on the relative depletion or enrichment of the natural radioelements. Most of the Th/U data are linearly correlated, grouped into two main categories, exhibiting Th/U ratios of 2 and 3. The theoretical value of the radioelement ratio Th/U can be calculated according to:

$$(Th/U) = \frac{T_{^{232}Th}}{T_{^{238}U}} \cdot \frac{f_{^{238}U}}{f_{^{232}Th}} \cdot \frac{M_{^{232}Th}}{M_{^{238}U}} \qquad (1)$$

where $T$ is the half-life, $f$ is the fractional atomic abundance, and $M$ is the atomic mass of the corresponding radioisotope indicated. Using values published in recent literature (Firestone et al., 1996-1999), the above formula yields the ratio Th/U $\cong$ 3.0, which is the typical value observed in normal continental crust.

The K/Th data (Fig. 4b) revealed less pronounced correlations, with typical ratios ranging between $1.5\times10^3$ and $3\times10^4$. For comparison, the corresponding Th/U and K/Th correlation plots from surface soils and rocks of sedimentary origin (regions R1, R2, R3, R5) are presented in Fig. 5. As can be seen, the Th/U ratio ranges from 1 to 6, while most of the K/Th data range between $1.5\times10^3$ and $3\times10^3$. Also, the Th/U and K/Th data for soils of ophiolitic origin (Fig. 4a,b) appear to be clustered within a narrow range due to the low content in these three radioelements. On the other hand, due to the very low concentration of Th and also of U in the ophiolite rocks, the K/Th ratios exhibit much higher values in ophiolites (Fig. 4b) compared to those of sedimentary origin (Fig. 5b). This result confirms previous indications based on averaged data extracted from the seven geological regions considered (Tzortzis and Tsertos, 2004). The observed features are in general agreement with the experimental results reported in a series of investigations in the Alps-Apeninnes transition by Pasquale et al. (2001) and Chiozzi et al. (2002), in the Pindos ophiolite by Valsami-Jones and Ragnarsdottir (1997), and in the Samail ophiolite by Chen and Pallister (1981).

Due to bleaching and eluviation processes during weathering of a rock, a loss in the radioelement concentration and, consequently, in the resulting natural radioactivity and dose rate is observed in the underlying surface soils. A measure, ε, for the total radioactivity loss can be calculated according to:

$$\varepsilon = \frac{\langle p \rangle}{\langle q \rangle} \quad (2)$$

where $p$ is the total radioactivity or, equivalently, the total dose rate in the soil, and $q$ the corresponding value in the rock. Since the dose rate on the soil surface and that of the rock surface are proportional to the sum of the weighted activity concentrations of

$^{232}$Th, $^{238}$U, and $^{40}$K of the soil and rock, one can use the known dose rate conversion factors of $^{232}$Th, $^{238}$U, and $^{40}$K (Tzortzis et al., 2003) to estimate ε:

$$\varepsilon = \frac{\langle ^{232}Th \rangle_{soil} \times 0.528 + \langle ^{238}U \rangle_{soil} \times 0.389 + \langle ^{40}K \rangle_{soil} \times 0.039}{\langle ^{232}Th \rangle_{rock} \times 0.528 + \langle ^{238}U \rangle_{rock} \times 0.389 + \langle ^{40}K \rangle_{rock} \times 0.039} \quad (3)$$

Using this relation and the measured average activity concentrations, shown in Table 1, a value of $\varepsilon \cong 0.46$ is calculated which is very close to the value (0.47) obtained from world data, recompiled recently by Minato (2002). This means that, on an average, nearly 50% of the radioelement content in the rocks of the Troodos ophiolite complex is depleted in the underlying surface soils. It is important to note here that a very similar value for ε is obtained, when the island's average concentrations for soils and rocks, shown in Table 2, are used in Eq. (3). Finally, this argumentation might be reversed to explain in a quite natural way the larger radioactivity levels observed in rocks (Tzortzis et al., 2003) compared to the corresponding values obtained from the underlying surface soils (Tzortzis et al., 2004), as being due to the effect of bleaching and eluviation during weathering of the rocks.

**Conclusions**

High-resolution γ–ray spectrometry is a powerful experimental tool in studying natural radioactivity and determining elemental concentrations in various surface soils and rock formations. For the predominant geological formations of the Troodos Ophiolitic Complex that were investigated appear generally to have lower radionuclide concentrations, as compared to those of sedimentary origin. The average values (A.M ± S.D.) derived are (0.24 ± 0.34) *ppm*, (0.10 ± 0.10) *ppm* and (0.21 ± 0.24) *%*, for Th, U, and K, respectively, in the soils, and (0.52 ± 0.17) *ppm*, (0.17 ± 0.11) *ppm* and (0.49 ± 0.87) *%* in the rocks. These values fall within the typical range of other ophiolites studied worldwide by other authors, revealing that ophiolites exhibit very low levels of natural radioactivity. Data on Th/U and K/Th correlations for soils of ophiolitic origin appear to be clustered within a narrow range due to the low content in these three radioelements. Due to the very low concentration of Th and also of U in the ophiolite rocks, the K/Th ratios exhibit much higher values in ophiolites compared to those of sedimentary origin. The measured Th/U ratio exhibits values between 2 and 4, whereas the K/Th ratio is highly variable ranging between $1.5 \times 10^3$ and $3.0 \times 10^4$. Finally, a radioactivity (radioelement) loss of nearly 50% is estimated in the underlying surface soils due to bleaching and eluviation during weathering of the rocks.

## Acknowledgements

This work is financially supported by the Cyprus Research Promotion Foundation (Grant No. 45/2001), and partially by the University of Cyprus.

**TABLE CAPTIONS**

**Table 1.** Summary of Arithmetic Mean (A.M.) and Standard Deviation (S.D.) of Th, U and K activity and elemental concentrations in samples from surface soils (Tzortzis et al., 2004) and main rock formations originated from the Troodos ophiolitic complex.

**Table 2.** Summary of Th, U, and K elemental concentrations reported from other known ophiolite complexes in neighbouring countries, together with the corresponding values obtained from the Troodos ophiolite. For comparison, the island's average values obtained from surface soils and main rock formations are also presented.

**Table 1.**

| Type | Number of Samples | Average Concentration (A.M ± S.D.) | | | | | |
|---|---|---|---|---|---|---|---|
| | | Activity concentration | | | Elemental concentration | | |
| | | $^{232}$Th (Bq kg$^{-1}$) | $^{238}$U (Bq kg$^{-1}$) | $^{40}$K (Bq kg$^{-1}$) | Th (ppm) $\times 10^{-1}$ | U (ppm) $\times 10^{-1}$ | K (%) $\times 10^{-1}$ |
| Soils | 59 | 1.0 ± 1.4 | 1.3 ± 1.3 | 65 ± 72 | 2.4 ± 3.4 | 1.0 ± 1.0 | 2.1 ± 2.4 |
| Rocks | 10 | 2.1 ± 0.7 | 2.1 ± 1.3 | 150 ± 260 | 5.2 ± 1.7 | 1.7 ± 1.1 | 4.9 ± 8.7 |

**Table 2.**

| Ophiolite, Region | Th (ppm) | U (ppm) | K (%) | Reference |
|---|---|---|---|---|
| Alps-Apennines, Italy | $<0.3 - 1.3$ | $<0.3 - 0.6$ | $<0.03 - 0.51$ | Chiozzi et al. (2002) |
| Pindos, Greece | $0.05 - 0.39$ | $0.05 - 0.31$ | – | Valsami-Jones and Ragnarsdottir (1997) |
| Samail, Oman | $0.05 - 0.67$ | $0.02 - 0.27$ | – | Chen and Pallister (1981) |
| Troodos, Cyprus | $<10^{-2} - 2.0$ | $<10^{-2} - 0.6$ | $<10^{-3} - 1.0$ | Present study |
| Cyprus average (soils) | $1.2 \pm 1.7$ | $0.6 \pm 0.7$ | $0.4 \pm 0.3$ | Tzortzis and Tsertos (2004) |
| Cyprus average (rocks) | $2.8 \pm 0.7$ | $1.3 \pm 0.3$ | $0.6 \pm 0.1$ | Tzortzis et al. (2003) |
| Worldwide average | 7.4 | 2.8 | 1.3 | UNSCEAR report (2000) |

# FIGURE CAPTIONS

**Figure 1**. Simplified geological map of Cyprus, indicating the Troodos Ophiolitic complex with its three geological regions discussed in this survey; the various rock formations appearing are described in the legend.

**Figure 2**. The stratigraphy of an Ophiolite model appears exactly the same to the layer structure of the ocean crust. Here, the various geological layers are drawn as a function of the ocean crust depth.

**Figure 3**. The Troodos ophiolite stratigraphy (R4, R6, and R7) along with the other geological layers appearing out of the ophiolite sequence (R1, R2, R3, and R5).

**Figure 4.** (a) Th versus U and (b) K versus Th concentration plots from surface soils and rocks of the Troodos ophiolite. Solid lines represent different values of ratios indicated.

**Figure 5.** The same as in Fig. 4, but for surface soils and rocks of sedimentary origin (R1, R2, R3, R5 regions shown in Fig. 3)

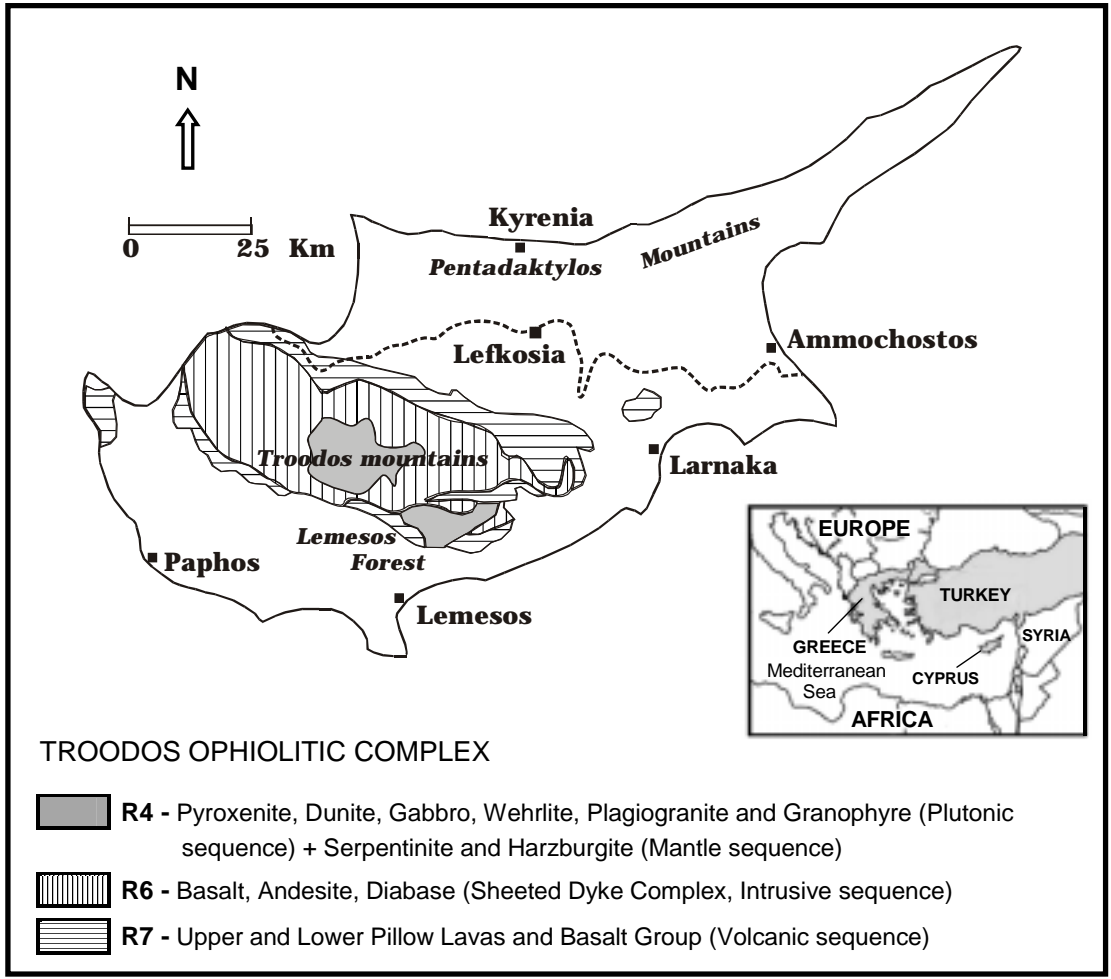

**Figure 1**.

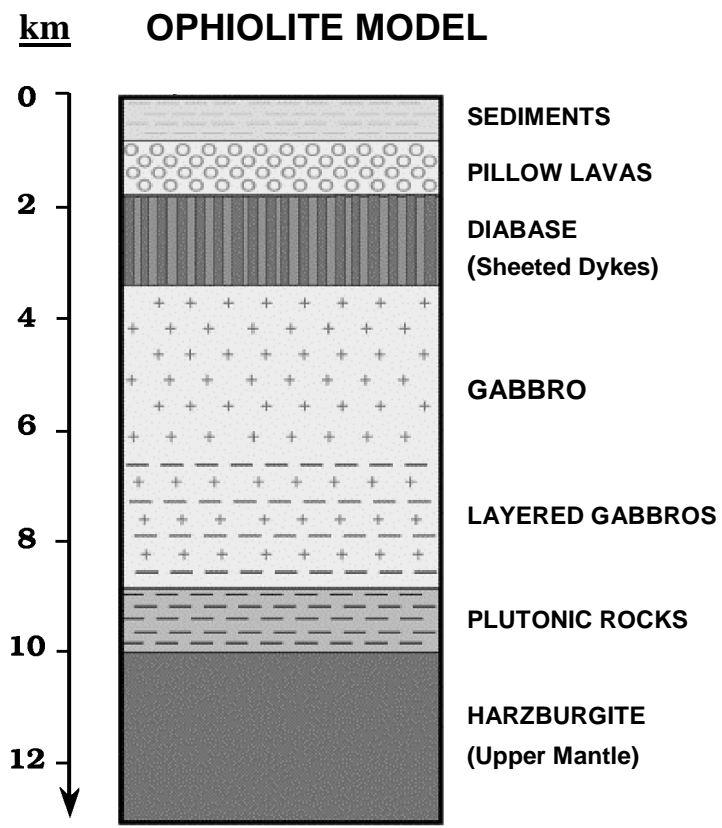

**Figure 2**.

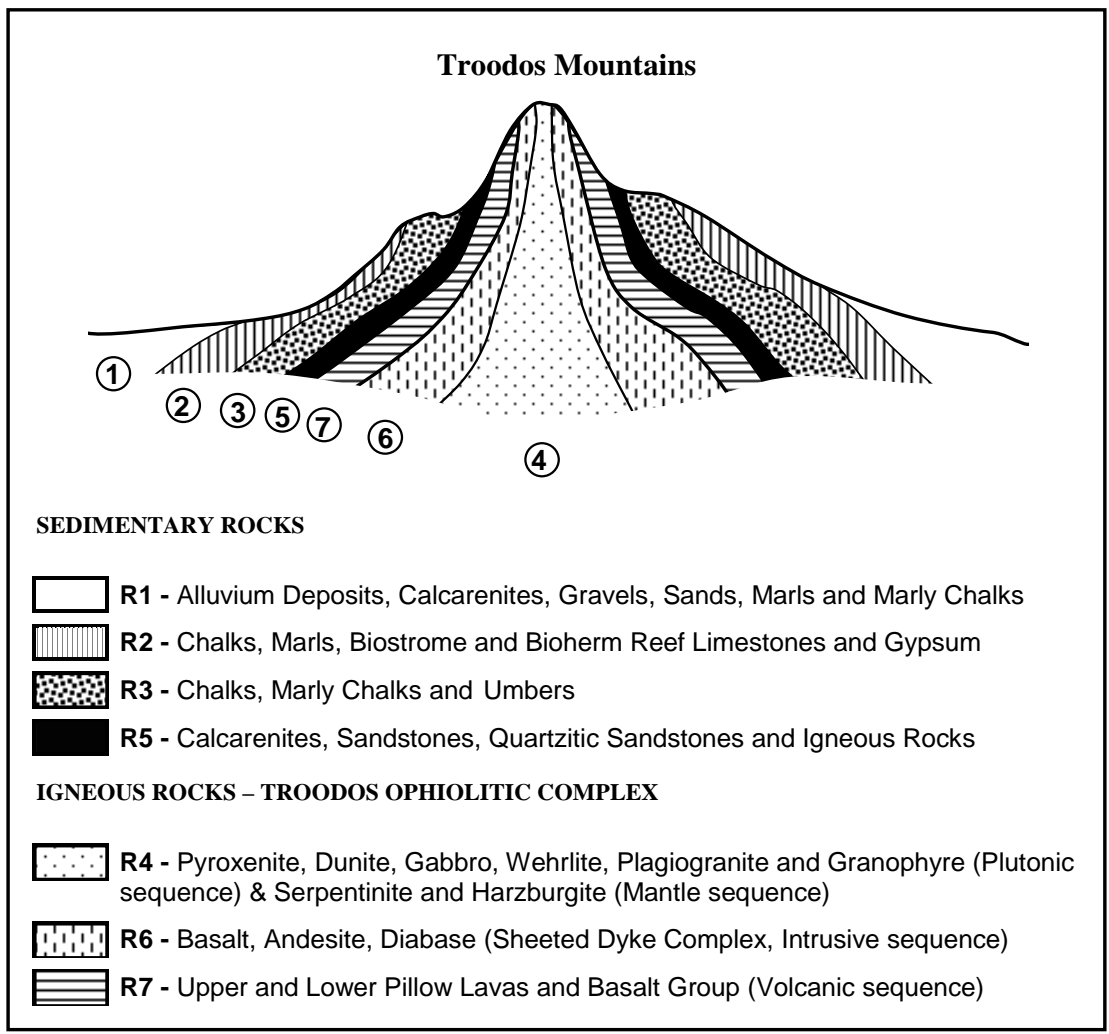

**Figure 3**.

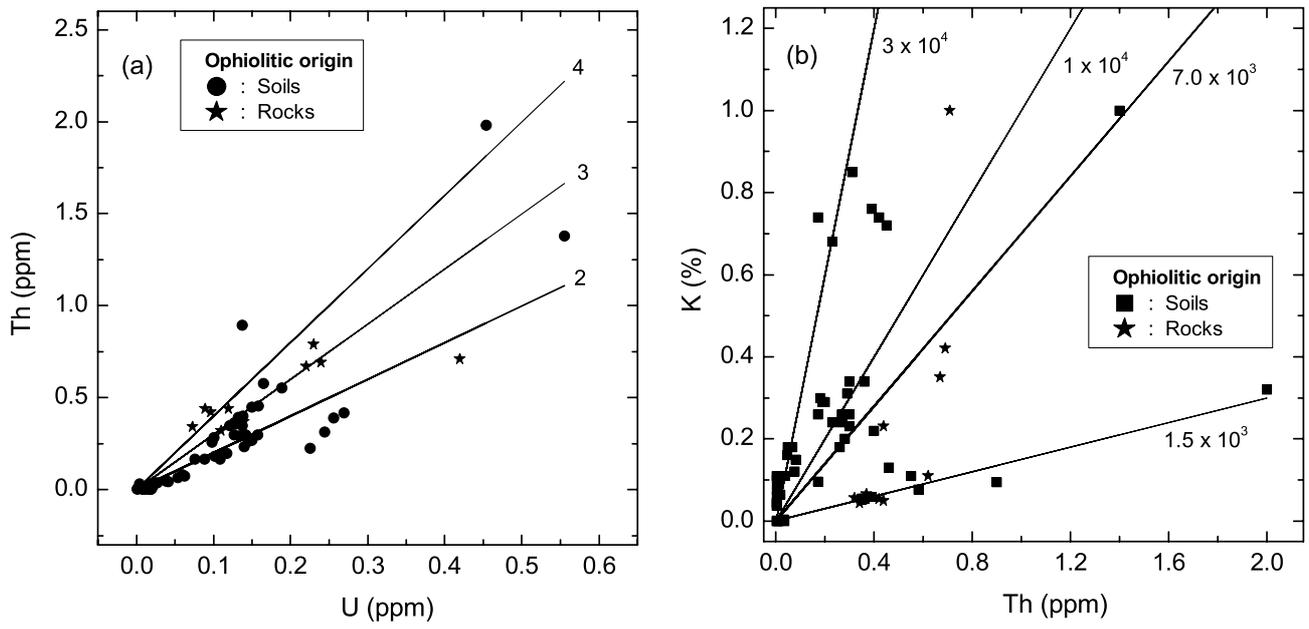

**Figure 4.**

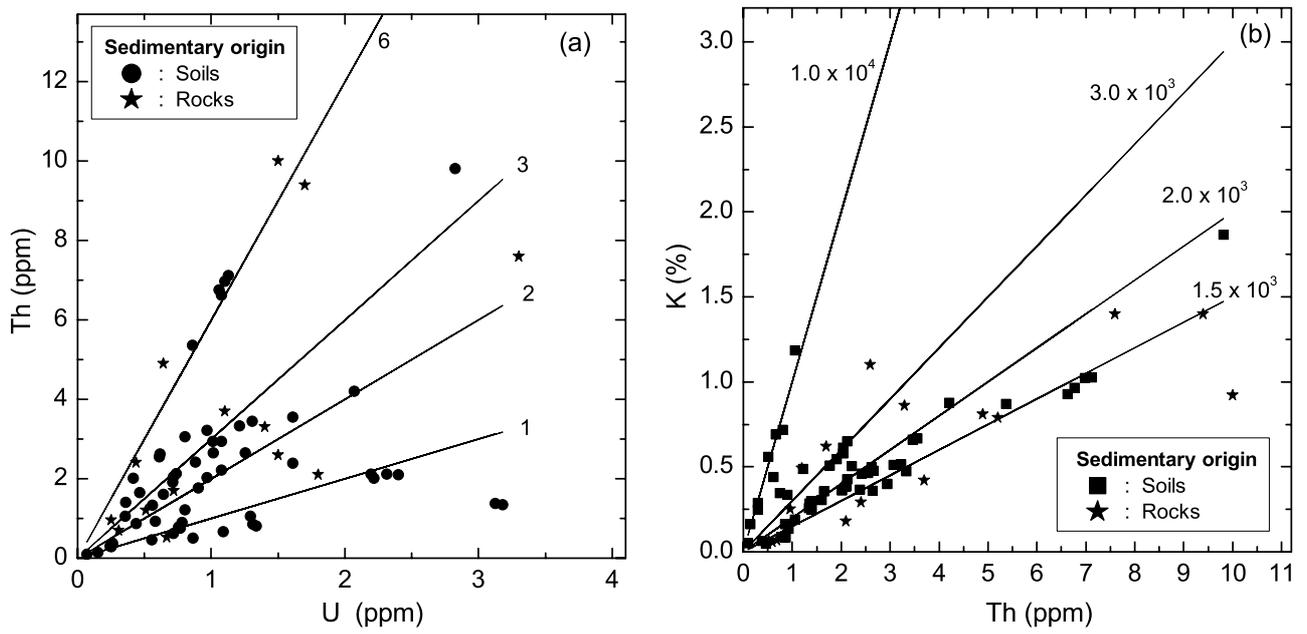

Figure 5.